\documentclass[12pt]{iopart}
\usepackage{iopams}
\usepackage{graphicx}
\usepackage{subfigure}

\newcommand{\eperp}{\mbox{$\epsilon_\perp$}}
\newcommand{\epar}{\mbox{$\epsilon_\parallel$}}
\newcommand{\ew}{\mbox{$\epsilon_w$}}

\DeclareFontFamily{U}{euc}{}
\DeclareFontShape{U}{euc}{m}{n}{<-6>eurm5<6-8>eurm7<8->eurm10}{}%
\DeclareSymbolFont{AMSc}{U}{euc}{m}{n} 
\DeclareMathSymbol{\micro}{\mathord}{AMSc}{"16}

\begin{document}

\title[Prolate-to-oblate shape transition of phospholipid
vesicles]{The prolate-to-oblate shape transition of phospholipid
vesicles in response to frequency variation of an AC electric field
can be explained by the dielectric anisotropy of a phospholipid
bilayer}

\author{Primo\v{z} Peterlin${}^1$, Sa\v{s}a Svetina${}^{1,2}$, and
Bo\v{s}tjan \v{Z}ek\v{s}${}^{1,2}$}

\address{${}^1$University of Ljubljana, Faculty of Medicine, Institute
of Biophysics, Lipi\v{c}eva 2, Ljubljana, Slovenia, ${}^2$Jo\v{z}ef Stefan
Institute, Jamova 39, Ljubljana, Slovenia}

\ead{primoz.peterlin@mf.uni-lj.si}

\date{\today}

\begin{abstract}
The external electric field deforms flaccid phospholipid vesicles into
spheroidal bodies, with the rotational axis aligned with its
direction. Deformation is frequency dependent: in the low frequency
range ($\sim 1$~kHz), the deformation is typically prolate, while
increasing the frequency to the 10~kHz range changes the deformation
to oblate. We attempt to explain this behaviour with a theoretical
model, based on the minimization of the total free energy of the
vesicle. The energy terms taken into account include the membrane
bending energy and the energy of the electric field. The latter is
calculated from the electric field via the Maxwell stress tensor,
where the membrane is modelled as anisotropic lossy
dielectric. Vesicle deformation in response to varying frequency is
calculated numerically. Using a series expansion, we also derive a
simplified expression for the deformation, which retains the frequency
dependence of the exact expression and may provide a better substitute
for the series expansion used by Winterhalter and Helfrich, which was
found to be valid only in the limit of low frequencies. The model with
the anisotropic membrane permittivity imposes two constraints on the
values of material constants: tangential component of dielectric
permittivity tensor of the phospholipid membrane must exceed its
radial component by approximately a factor of 3; and the membrane
conductivity has to be relatively high, approximately one tenth of the
conductivity of the external aqueous medium.
\end{abstract}

\pacs{87.16.Dg, 87.50.Rr, 77.84.-s, 41.20.Cv}
\submitto{\JPCM}


\section{Introduction}

Rapid development in biotechnology induced a growing interest in the
influence of the AC electric field on biological cells, phospholipid
vesicles and colloidal particles
\cite{Zimmermann:Electromanipulation}, which instigated theoretical
investigations in this area
\cite{Jones:Electromech,Foster:1996}. Among the systems studied,
phospholipid vesicles are often chosen as the model system of choice
due to their well-defined structure.

In the late 1950s, Schwan \cite{Schwan:1957} began a series of studies
of the influence of the electric field on biological cells, modelling
them as simple geometric shells with given electric properties. In the
early 1970s, Helfrich included the then newly developed elastic theory
for lipid bilayers \cite{helfrich:1973} into the treatment of the
effect of the external electric field on phospholipid vesicle shape
\cite{helfrich:1974A}. The first experiments with phospholipid
vesicles in an electric field which followed several years later
\cite{harbich:1979} seemed to confirm the claim from the
previous theoretical analysis \cite{helfrich:1974A}: a 2~kHz external
AC electric field deforms flaccid phospholipid vesicles into
\textit{prolate} spheroids. Winterhalter and
Helfrich~\cite{winterhalter:1988} extended the earlier theoretical
treatment \cite{helfrich:1974A}; their model allows for a finite
electrical resistance of the bilayer, an AC electric field, and
includes Maxwell stresses inside the membrane.  The authors treated
the spherical vesicle as a lossy dielectric immersed into a medium
which was also treated as a lossy dielectric. They examined the case
$\sigma_m/\sigma_w \ll \epsilon_m/\ew$, where $\sigma_m$, $\sigma_w$,
$\epsilon_m$ and $\ew$ are the conductivities and permittivities of
the membrane and the aqueous medium, respectively. Employing series
expansions, they obtained a simple expression for vesicle
deformation and predicted a conducting regime of vesicle behaviour at
low frequencies ($\omega \ll \sigma_m/\epsilon_m$), where the field
does not penetrate the vesicle interior, a dielectric regime at higher
frequencies ($\omega \gg \sigma_w/\ew$), where the field penetrates
the vesicle interior, and an intermediate regime in between the
two. In all three regimes, they obtained a prolate vesicle shape.

\begin{figure}
  \subfigure[]{\includegraphics[width=0.5\linewidth]{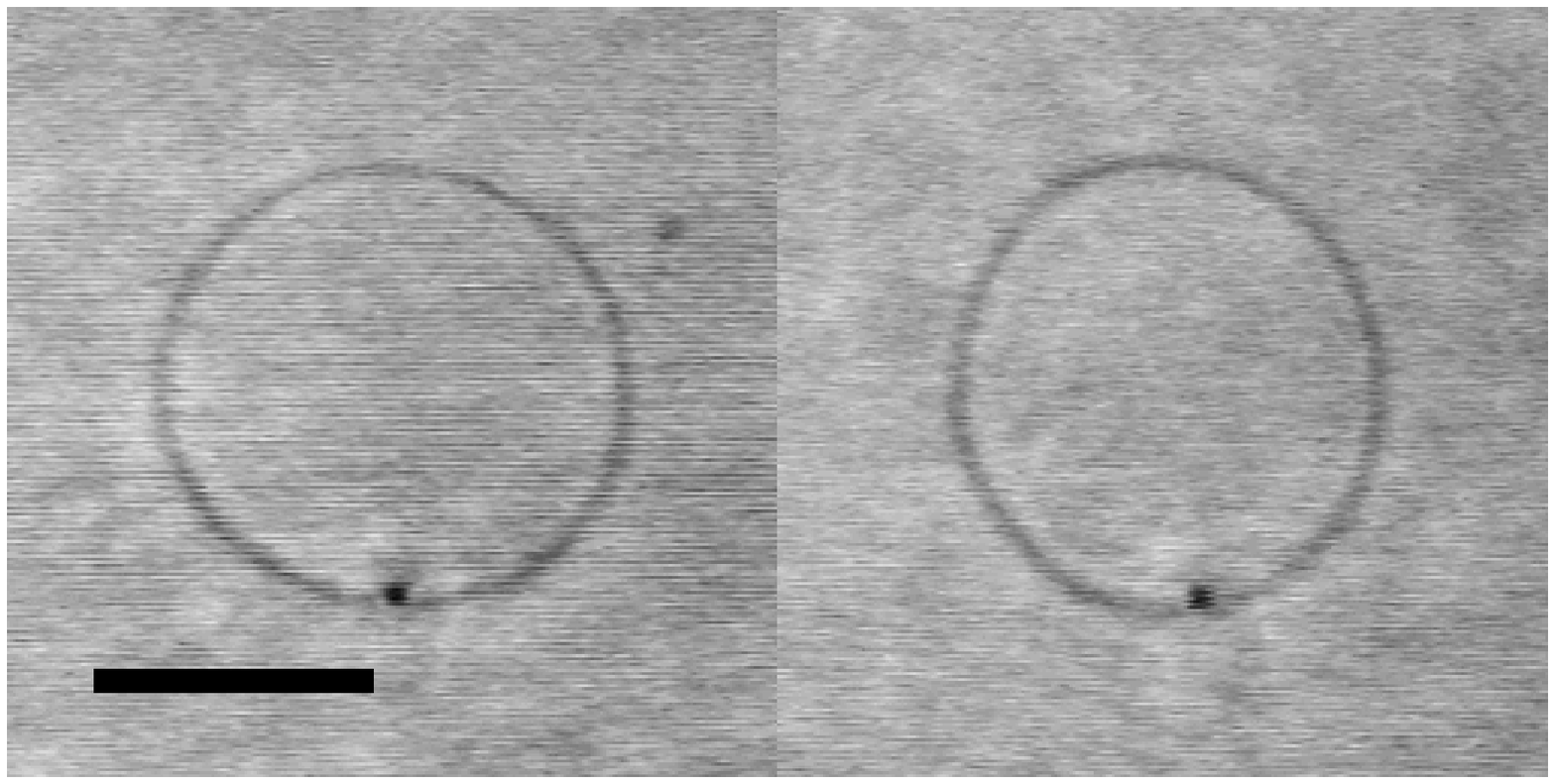}}\hfil
  \subfigure[]{\includegraphics[width=0.45\linewidth]{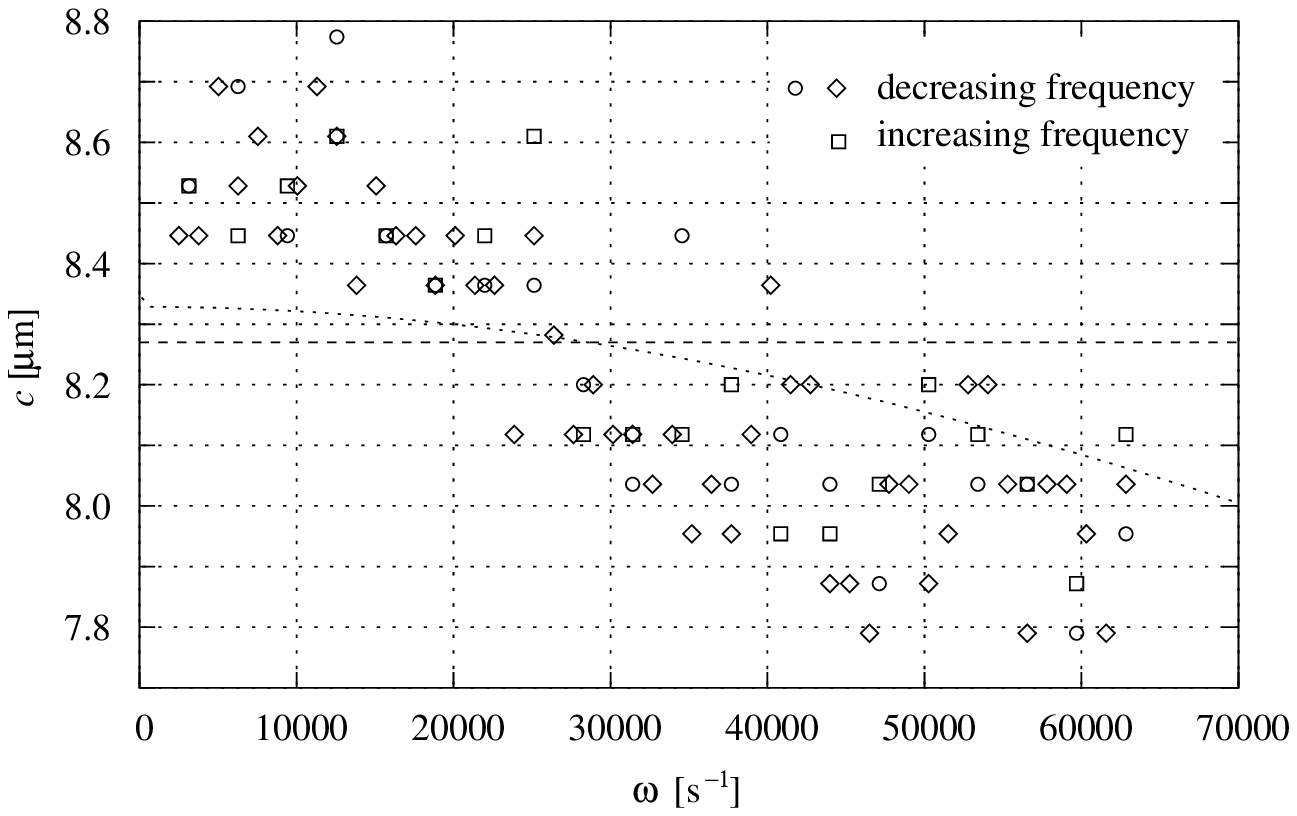}}
  \caption{(a) Giant phospholipid vesicle is deformed into a prolate
  shape (left) at an applied AC field of 1~kHz, and into an oblate
  shape at 10~kHz (right). The conductivity of the aqueous medium was
  1.3~$\micro$S/cm. An electric field of $\approx 2\cdot 10^4$~V/m was
  applied in the horizontal direction. The bar represents
  10~$\micro$m. (b) Deformation of the same vesicle, expressed as the
  value of the semiaxis $c$ (aligned with the direction of the
  electric field) as a function of the circular frequency $\omega$ of
  the external AC electric field. The dashed line corresponds to the
  value at which the vesicle is spherical; greater values of $c$
  correspond to prolate shapes, smaller to oblate. The dotted line is
  discussed further in the text. The three consecutive runs yield
  $2394\pm75$, $2427\pm75$ and $2371\pm87$ $\micro$m${}^3$, as
  estimates for the vesicle volume, respectively, which indicates that
  the vesicle volume did not change significantly during the
  experiment. Details on experimental setup are given in
  \cite{Peterlin:2000a}.}
  \label{fig:vesicle}
  \label{fig:popc157sigmoid}
\end{figure}

In contrast to that, experiments with varying frequency
\cite{mitov:1993,peterlin:1993} have demonstrated that a vesicle
undergoes a \textit{prolate-to-oblate shape change} when the frequency
of the applied field is increased from the 1~kHz range into the 10-kHz
range (figure~\ref{fig:vesicle}a). Figure~\ref{fig:popc157sigmoid}b
shows the dependence of the ratio of vesicle semiaxes $c/a$ for the
shown vesicle on the frequency of the applied electric field, with $c$
being the semiaxis parallel to the field, and $a$ the semiaxis
perpendicular to it.

In this paper, we take the work of Winterhalter and Helfrich as our
starting point and proceed to extend it by allowing for an anisotropy
of the permittivity of the phospholipid bilayer. Since the model of
Winterhalter and Helfrich departs from the experimental data in the
high-frequency region by failing to predict oblate shapes, we sought a
modification that would affect its behaviour in the high-frequency
range, where the behaviour of the system is governed solely by the
permittivities.

Recently, several papers have been published which take into account
the anisotropy of the membrane dielectric permittivity and the
membrane electric conductivity, which arise due to adsorbed
hydrophobic ions on the membrane \cite{Sukhorukov:2001,Ko:2004}.
Ambj\"ornsson and Mukhopadhyay \cite{Ambjornsson:2003} have provided a
solution for the electric potential for a general ellipsoid coated
with a dielectrically anisotropic coating; the same group also
examined a possible mechanism for the dielectric anisotropy in the
membrane at frequencies where molecular resonances are important
\cite{Ambjornsson:2004}. Simeonova and Gimsa \cite{Simeonova:2005}
have extended the description of the phospholipid vesicle membrane to
a three-layer shell to take into account the fact that the dielectric
anisotropy occurs only in the phospholipid headgroup layer, while the
middle layer of hydrocarbon chains is mainly isotropic. It also needs
to be noted that transitions between prolate and oblate shapes in
response to a frequency variation of the applied electric field have
been examined theoretically for cases where the electrical properties
of the aqueous solution inside the vesicle differ from those in the
vesicle exterior \cite{Hyuga:1991C,Hyuga:1993}, and recently, an
experimental morphological phase diagram has also been obtained
\cite{Aranda:2006}. A review of the studies of the influence of the
electric field on phospholipid vesicles was recently summarized in a
paper by Dimova and coworkers \cite{Dimova:2006}.

While the above papers provide an analytical expression for the
electric potential \cite{Ambjornsson:2003} or a thorough numerical
analysis of the Clausius-Mossotti factor
\cite{Sukhorukov:2001,Ko:2004,Simeonova:2005}, they do not focus on an
analysis of parameters at which either prolate or oblate deformation
is possible in the high- and low-frequency limit, which is the aim of
this paper.

The paper is organized as follows. In section~2 we first develop an
extension of the model of Winterhalter and Helfrich by taking into
account the anisotropy of membrane permittivity. In section~3 we show
that the model presented here exhibits frequency-dependent
prolate-to-oblate shape transition of a phospholipid vesicle similar
to the one observed in the experiment. In the expansion valid for
large vesicles, we derive the neccessary conditions for a prolate or
oblate deformation in both the high- and low-frequency limit, and
analyse the dependence of the prolate-to-oblate shape transition
frequency on material constants. Finally, in section~4, we discuss
some of the issues which arise from the results.

\section{Theoretical framework}

The equilibrium vesicle shape in an external electric field is
calculated as the shape with the minimal total free energy, consisting
of the vesicle bending energy and the energy due to the electric
field. The calculation was conducted for a vesicle with constant
volume, while the necessary area increase for the deformation stems
from the flattening of thermal fluctuations of the vesicle by the
electric field \cite{Kummrow:1991,Niggemann:1995}. The analytical
expression for the membrane bending energy is well known
\cite{helfrich:1973}, and the change of the energy due to the electric
field is calculated as the work done by the force of the electric
field while deforming the vesicle \cite{winterhalter:1988}. In order
to evaluate it, we first have to calculate the electric field. The
present treatment is limited to small deviations of vesicle shape from
the sphere, thus the electric field in the presence of a spherical
shell is computed. Both the aqueous solution inside and outside the
vesicle and the vesicle membrane are treated as lossy dielectrics. The
aqueous solutions inside and outside the vesicle have identical
electrical properties. Dielectric permittivity of the membrane is
treated as anisotropic, with the component aligned with the normal to
the membrane differing from the components perpendicular to this
direction. Both the dielectric permittivity and the electric
conductivity of the aqueous solution as well as the electric
conductivity of the membrane are treated as scalars.

The applied electric field $\mathbf{E}$ introduces a single distinct
axis into the system, so the treatment can be limited to axially
symmetric shapes. This eliminates the dependence on the longitude
angle $\phi$, if the polar ($z$) axis is chosen parallel to the applied
field. Thus, the vesicle surface can be parametrized as
\begin{equation}
  r(\theta) = s_0 + s(\theta)\;,
\end{equation}
where $|s(\theta)| \ll s_0$, $s_0$ denoting the deformation
independent of the polar angle $\theta$. In an absence of deformation
($|s(\theta)|=0$), $s_0$ equals the radius of the undeformed sphere
$r_0$. The deformation is independent of the sign of the electric
field, thus it is proportional to $E^2$ in the lowest order. As the
field itself is proportional to $\cos\theta$, one can expect a
deformation coupled with a field to be proportional to
$\cos^2\theta$. In terms of expansion into spherical harmonics, this
limits us to a sum of even terms. Retaining only the terms
proportional to $E^2$ or lower, a quadrupolar term remains, where
$s(\theta)$ equals the second Legendre polynomial: $s(\theta) =
\frac{1}{2} s_2 (3\cos^2 \theta - 1)$, with $s_2$ being a measure for
the extent of deformation. Positive values of $s_2$ indicate prolate
deformation, negative oblate.

If the membrane area is to be locally conserved, a quadrupolar
displacement $\delta r_r$ in a radial direction must be accompanied by
a tangential displacement $\delta r_\theta$ \cite{winterhalter:1988}:
\numparts
\begin{eqnarray} 
  \delta r_r & = & \frac{1}{2} (3 \cos^2 \theta - 1) \,s_2
  \label{eq:displacement-r} \\ 
  \delta r_\theta & = & -\cos\theta
  \sin\theta \: s_2 \; .
  \label{eq:displacement-th}
\end{eqnarray}
\endnumparts 
A requirement for a local area conservation assures that the membrane
stretching is independent of polar angle $\theta$.  The total membrane
area expansion is determined by the relationship between $s_0$ and
$r_0$. Taking into account the requirement for a constant vesicle
volume, the following relationship for $s_0$ is obtained:
\begin{equation}
  s_0 = r_0 - \frac{s_2^2}{5 r_0} \; .
  \label{eq:volume-renorm}
\end{equation}
The correction for a constant volume (\ref{eq:volume-renorm}) contains
a higher term in the powers of $s_2$ and thus does not affect in the
lowest term either the bending energy or the energy due to the
electric field, yielding $s_0=r_0$ an adequate approximation.

For small deformations, quadrupolar deformation only induces small
perturbative changes in the vesicle bending energy and the energy due
to the electric field.
\begin{eqnarray}
  G_\mathrm{bend}(s_2) &\approx G_\mathrm{bend}(s_2=0) + \frac{1}{2}
  \left.\frac{\partial^2 G_\mathrm{bend}}{\partial
  s_2^2}\right|_{s_2=0}\!\! s_2^2
  \label{eq:Gbend-perturb} \\
  G_\mathrm{field}(s_2) &\approx G_\mathrm{field}(s_2=0) + 
  \left.\frac{\partial G_\mathrm{field}}{\partial
  s_2}\right|_{s_2=0}\!\! s_2
  \label{eq:Gfield-perturb}
\end{eqnarray}
The total energy of the vesicle formally depends on two parameters,
$r_0$ and $s_2$. The constraint requiring a constant vesicle volume
however eliminates one degree of freedom, yielding
(\ref{eq:Gbend-perturb}--\ref{eq:Gfield-perturb}). It is also worth
noting that the fact that neither prolate ($s_2> 0$) nor oblate shapes
($s_2 < 0$) have a bending energy lower than those of a sphere means
that the expansion for the bending energy (\ref{eq:Gbend-perturb})
contains no linear term.

Equilibrium vesicle deformation, expressed in terms of $s_2$, can then
be calculated by minimizing the total free energy over $s_2$:
\begin{equation}
  \frac{d}{d s_2} \left( G_\mathrm{bend} +
  G_\mathrm{field} \right) = 
  \left.\frac{\partial^2 G_\mathrm{bend}}{\partial
  s_2^2}\right|_{s_2=0}\!\! s_2 + \left.\frac{\partial
  G_\mathrm{field}}{\partial s_2}\right|_{s_2=0} = 
  0 \; .
  \label{eq:equilibrium-requirement}
\end{equation}
Equation~(\ref{eq:equilibrium-requirement}) gives the extent of
deformation ($s_2$) at given conditions.

It is now our task to write an expression for the vesicle
bending energy (\ref{eq:Gbend-perturb}), which can be written as
\cite{helfrich:1973}:
\begin{equation}
  G_\mathrm{bend} = \frac{1}{2}k_c \oint \left( c_1 + c_2 \right)^2 dA
  \; .
\end{equation}
Here, $k_c$ is the bending elastic modulus of the membrane, while
$c_1$ and $c_2$ are the principal curvatures of the membrane. The
spontaneous curvature $c_0$ has been omitted, because it vanishes for
a bilayer composed of two equal layers. The integration is conducted
over the total membrane area $A$ of a quadrupolarly deformed
vesicle. Up to quadratic order terms in $s_2$, the total bending
energy of a nearly spherical vesicle can be written as
\cite{helfrich:1973}:
\begin{equation}
  G_\mathrm{bend} = 8\pi k_c + \frac{48\pi}{5} k_c
  \left(\frac{s_2}{r_0}\right)^2 \, .
  \label{eq:Gbend}
\end{equation}
This can be readily interpreted as the bending energy of a sphere plus
an addition due to the quadrupolar deformation. 

\begin{figure}
  \includegraphics[width=8cm]{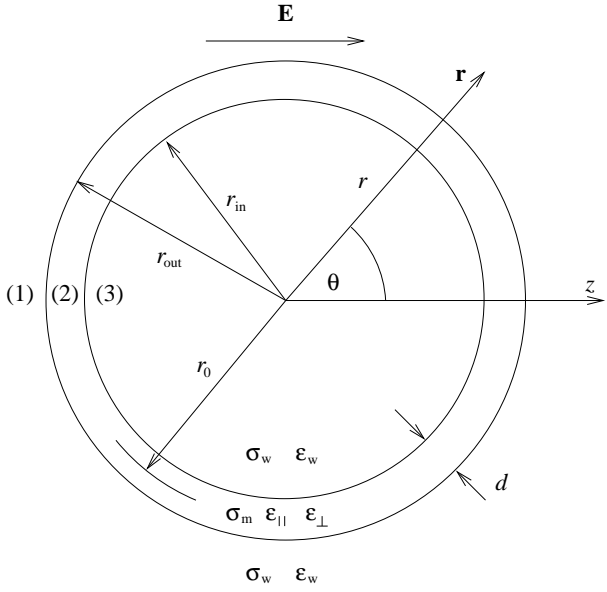}
  \caption{A vesicle is modelled in spherical coordinates $(r,\theta)$
    as a spherical shell with the external radius $r_\mathrm{out} =
    r_0 + d/2$ and the internal radius $r_\mathrm{in} = r_0 - d/2$
    ($d\ll r_\mathrm{in}$ being the membrane thickness) exposed to an
    external electric field $\mathbf{E}$. The conductivity and
    permittivity of the aqueous solution inside and outside the
    membrane are denoted by $\sigma_w$ and $\epsilon_w$, and the
    membrane conductivity is denoted by $\sigma_m$. Membrane
    permittivity is in general anisotropic, hence $\epar$, $\eperp$;
    their orientation is defined in the text
    (equation~\protect\ref{eq:permittivity-tensor}). The three media,
    vesicle exterior, vesicle membrane and vesicle interior are
    denoted by (1), (2), and (3), respectively.}
  \label{fig:spherical-shell}
\end{figure}

The free energy term arising from the electric field
(\ref{eq:Gfield-perturb}) is calculated as the work done by the forces
of the electric field during the deformation of the vesicle. In order
to evaluate it, we first have to calculate the electric field around a
vesicle (figure~\ref{fig:spherical-shell}). Gauss' law
$\nabla\cdot\mathbf{D}=0$ together with the requirement for
an irrotational electric field $\nabla\times\mathbf{E}=0$, which stems
from Faraday's law for electromagnetic induction, yields the
Laplace equation for the electric potential $U$ in a medium with
homogenous dielectric permittivity \cite{Landau:Electrodynamics} -- in
our case the aqueous medium in the vesicle interior and the external
aqueous medium (respectively denoted by (3) and (1) in
figure~\ref{fig:spherical-shell}). In the anisotropic case, however,
Gauss' law does not lead to the Laplace equation for the electric
potential $U$, $\nabla\cdot\mathbf{D} =
\nabla\cdot\underline{\epsilon}\,\mathbf{E} =
-\nabla\cdot(\underline{\epsilon}\nabla U)$.  Here, $\mathbf{D}$ is
the displacement field, $\mathbf{E}$ is the electric field and
$\underline{\epsilon}$ is the permittivity tensor.

The boundary conditions on both the outer and the inner membrane-water
boundary impose $\nabla\times\mathbf{E} = 0$, thus requiring the
continuity of the tangential component of the electric field, and that
the total surface charge density, including both free charge and
displacement charge, must vanish. In the spherical geometry, this
yields the equations
\begin{eqnarray}
  E_\theta^{(1)} &= E_\theta^{(2)} \; , \label{eq:boundarycond-tangential} \\
  (\epsilon_w - i \omega\sigma_w) E_r^{(1)} &= (\epar - i
  \omega\sigma_m) E_r^{(2)} \; , \label{eq:boundarycond-radial}
\end{eqnarray}
for the (1--2) interface ($r=r_\mathrm{out}$,
figure~\ref{fig:spherical-shell}) and similarly for the (2--3)
interface ($r=r_\mathrm{in}$). Apart from
(\ref{eq:boundarycond-tangential}--\ref{eq:boundarycond-radial}), the
system is constrained by two additional conditions: one requiring that
the electric field far from the vesicle is unperturbed, and the other
requiring that the electric field is finite inside the vesicle.

In (\ref{eq:boundarycond-tangential},\ref{eq:boundarycond-radial}) we
introduced the notation for the dielectric permittivities and electric
conductivities in the three media: $\epsilon^{(1)} = \epsilon^{(3)} =
\epsilon_w$, $\sigma^{(1)} = \sigma^{(3)} = \sigma_w$, amd
$\sigma^{(2)} = \sigma_m$. The membrane permittivity is treated as
anisotropic. Due to its structure, electric properties in the
direction along the long axis of phospholipid molecules (or normal to
the vesicle membrane) are expected to be markedly different from the
properties in the direction perpendicular to it. The phospholipid
molecule is not axially symmetric per se, but due to the unordered
liquid-like structure of the bilayer we can approximate it as such on
the timescale of interest to our problem. The two directions
perpendicular to the normal on the membrane can thus be treated as
being equal. The component along the normal to the membrane is denoted
by $\epar$ and the permittivity in the direction perpendicular to it
by $\eperp$. Local permittivity in a given point on the membrane is
thus equal to $\epar$ in the normal direction and $\eperp$ in the
tangential direction. Permittivity tensor is locally defined as:
\begin{equation}
  \underline{\epsilon} = \left[
    \begin{array}{ccc}
      \epar 	& 0 	 & 0 \\
      0 	& \eperp & 0 \\
      0	& 0	 & \eperp 
    \end{array} \right] \; .
  \label{eq:permittivity-tensor}
\end{equation}

For a spherical shell, local coordinates conveniently coincide with
the spherical coordinates. Gauss' law
$\nabla\cdot(\underline{\epsilon}\nabla U)=0$ for the treated geometry
is most conveniently written in spherical coordinates,
\begin{equation}
  \epsilon_\parallel \frac{1}{r^2} \frac{\partial}{\partial r}
  \left( r^2 \frac{\partial U}{\partial r} \right) +
  \epsilon_\perp \frac{1}{r^2
    \sin\theta}\frac{\partial}{\partial \theta}
  \left( \sin\theta \frac{\partial U}{\partial\theta}\right) = 0 \; .
\end{equation}
In spherical coordinates, the equation can be separated into the
radial and the angular parts, $U(r,\theta) = R(r)\Theta(\theta)$ (it
has been already taken into account that the problem is axially
symmetrical and thus independent of $\phi$), yielding two separate
equations for the radial and the angular part. For the angular part, a
Legendre differential equation is obtained. The equation obtained for
the radial part differs from the spherical Bessel equation by
involving a factor $\eperp/\epar$ in the second term:
\begin{equation}
  \frac{1}{r^2} \frac{d}{dr} \left( r^2 \frac{dR}{dr} \right) -
  \frac{\eperp}{\epar} \frac{l(l + 1)}{r^2} R = 0 \; .
  \label{eq:radial-anisotrop}
\end{equation}
It is worth noting that (\ref{eq:radial-anisotrop}) represents a
special case of Heun's equation, which arises from solving the case of
a general ellipsoid
\cite{Ambjornsson:2003}. Equation~(\ref{eq:radial-anisotrop}) is
solved by a linear combination
\[
R_l(r) = C_1 r^{\frac{1}{2}(-1-\sqrt{1 +
4l\epsilon_\perp/\epsilon_\parallel +
4l^2\epsilon_\perp/\epsilon_\parallel})} + C_2
r^{\frac{1}{2}(-1+\sqrt{1 + 4l\epsilon_\perp/\epsilon_\parallel +
4l^2\epsilon_\perp/\epsilon_\parallel})} \; .
\]
The symmetry requirements of our problem limit us to the case $l=1$.
Thus, the ansatz for the electric potential that satisfies Gauss'
law for the anisotropic case is
\begin{equation}
U^{(2)} = \frac{1}{2} \biggl[ \left( a^{(2)} r^{(\alpha-1)/2} + b^{(2)}
  r^{(-\alpha-1)/2} \right) \cos\theta\, e^{-i \omega t} +
  \textrm{C.C.} \biggr] \; .
\label{eq:ansatz-anisotropic}
\end{equation}
A shorthand notation
$\alpha=\sqrt{1+8\epsilon_\perp/\epsilon_\parallel}$ has been
introduced along the way. The isotropic case corresponds to
$\alpha=3$, which simplifies (\ref{eq:ansatz-anisotropic}) to a known
form
\begin{equation}
  U^{(k)} = \frac{1}{2} \left[ \left( a^{(k)} r + \frac{b^{(k)}}{r^2}
    \right) \cos\theta\, e^{-i \omega t} + \textrm{C.C.} \right]\; ,
    \quad k=1,3 \; .
\end{equation}
The coefficients $a^{(k)}$, $b^{(k)}$; $k=1,2,3$, can in general be
complex to allow for a phase shift, and are determined from the
boundary conditions.

The boundary conditions for the unperturbed field far away from the
vesicle and the finite field inside the vesicle immediately yield two
coefficients:
\begin{eqnarray}
a^{(1)} & =  -E_0 \\
b^{(3)} & =  0
\end{eqnarray}
The remaining four coefficients are determined by the four equations
specifying boundary conditions (\ref{eq:boundarycond-tangential},
\ref{eq:boundarycond-radial}):
\numparts
\begin{eqnarray}
\fl a^{(2)} &= \frac{-6 E_0 r_\mathrm{out}^{(3-\alpha)/2}(2+\beta+\alpha\beta)}
	{8 + (2+6\alpha)\beta + (\alpha^2-1)\beta^2
	- \left[8 + (2-6\alpha)\beta + (\alpha^2-1)\beta^2\right]\gamma^\alpha} 
	\label{eq:a2-anisotrop} \\
\fl a^{(3)} &= \frac{-12 E_0 \alpha\beta\gamma^{(\alpha-3)/2}}
	{8 + (2+6\alpha)\beta + (\alpha^2-1)\beta^2
	- \left[8 + (2-6\alpha)\beta + (\alpha^2-1)\beta^2\right]\gamma^\alpha} \\
\fl b^{(1)} &= \frac{E_0 r^3 [(\alpha-1)\beta-2]
		(2+\beta+\alpha\beta)(1-\gamma^\alpha)}
	{8 + (2+6\alpha)\beta + (\alpha^2-1)\beta^2
	- \left[8 + (2-6\alpha)\beta + (\alpha^2-1)\beta^2\right]\gamma^\alpha} \\
\fl b^{(2)} &= \frac{-6 E_0 r_\mathrm{out}^{(3+\alpha)/2} [(\alpha-1)\beta-2] 
                \gamma^\alpha}
	{8 + (2+6\alpha)\beta + (\alpha^2-1)\beta^2
	- \left[8 + (2-6\alpha)\beta + (\alpha^2-1)\beta^2\right]\gamma^\alpha} 
	\label{eq:b2-anisotrop}
\end{eqnarray}
\endnumparts
Consistently with \cite{winterhalter:1988}, two more shorthand
notations have been introduced:
\begin{eqnarray}
  \beta &= \frac{\sigma_m - i \omega\epar}{\sigma_w - i \omega
	\epsilon_w} \label{eq:beta} \; , \\
  \gamma &= \frac{r_\mathrm{in}}{r_\mathrm{out}} \; .\label{eq:gamma}
\end{eqnarray}

The surface density of the force exerted on the boundary of
dielectrics by the electric field is equal to the scalar product of
the Maxwell stress tensor and the normal vector to the membrane,
\numparts
\begin{eqnarray}
\mathbf{f}_\mathrm{out} &= (\underline{T}^{(1)} - \underline{T}^{(2)})
\mathbf{e}_r \label{eq:fout} \; , \\
\mathbf{f}_\mathrm{in} &= (\underline{T}^{(2)} - \underline{T}^{(3)})
\mathbf{e}_r \; .
\label{eq:fin}
\end{eqnarray}
\endnumparts
The force vanishes in a homogeneous medium, but can in general be
non-zero on the boundaries of media with different electrical
properties. The Maxwell stress tensor is defined as
\begin{eqnarray}
  \underline{T} & = {\bf D}\otimes{\bf E} - \frac{1}{2}({\bf
    D}\cdot{\bf E})\underline{I} \;,
\end{eqnarray}
where $\underline{I}$ denotes the identity matrix.

Unlike in the case of the bending energy term in the total free energy
(\ref{eq:Gbend}), for which an analytical expression was obtained, an
approach where energy difference is computed is employed here. $\delta
G_\mathrm{field}$ denotes a small change in the energy due to the
electric field, when a sphere ($s_2=0$) is perturbed by a small
deformation change $\delta s_2$. This energy difference is calculated
as the work done by the forces of electric field during the
displacement of membrane elements $\delta\mathbf{r}$
(equations~\ref{eq:displacement-r},\ref{eq:displacement-th}),
integrated over the entire membrane area \cite{winterhalter:1988}:
\begin{equation}
\delta G_\mathrm{field} = 
- \oint (\mathbf{f}_\mathrm{out}\cdot\delta\mathbf{r})\,dA_\mathrm{out} 
- \oint (\mathbf{f}_\mathrm{in}\cdot\delta\mathbf{r})\,dA_\mathrm{in} \; .
\label{eq:Gfield-general}
\end{equation}
The integration is conducted over a sphere, which is consistent with
the limit of small deformations ($s_2 \ll r_0$).

Substituting the coefficients
(\ref{eq:a2-anisotrop}--\ref{eq:b2-anisotrop}) into
(\ref{eq:Gfield-general}) yields a lengthy expression for
$\delta G_\mathrm{field}$, which will not be reproduced here. After
substituting the expresson (\ref{eq:beta}) for $\beta$ into it, one
obtains an expression of the form $(A + B\omega^2 + C\omega^4)/(D +
E\omega^2 + F\omega^4)$, which can be written as a sum of two
dispersion terms:
\begin{equation}
\frac{dG_\mathrm{field}}{ds_2} \rightarrow \frac{\delta
  G_\mathrm{field}}{s_2} = -\frac{6\pi}{5} E_0^2 \epsilon_w
r_\mathrm{out}^2 \left( \xi_\infty + \frac{\xi_1}{1 + \omega^2
  \tau_1^2} + \frac{\xi_2}{1 + \omega^2 \tau_2^2} \right) \; ,
\label{eq:Gfield-omega}
\end{equation}
where the coefficients $\xi_\infty$, $\xi_1$, $\xi_2$, $\tau_1$ and
$\tau_2$ are rather lengthy expressions involving five different
material constants: $\epar$, $\eperp$, $\epsilon_w$, $\sigma_m$,
$\sigma_w$, the membrane thickness $d$ and the vesicle radius
$r_0$. It turns out, however, that the actual number of independent
parameters is lower. By introducing a dimensionless frequency
$\omega/(\sigma_w/\ew)$, one can reduce the frequency dependence of
(\ref{eq:Gfield-omega}) to only four parameters. Here, we have chosen
them to be $\gamma$, $\epar/\epsilon_w$, $\Delta\epsilon/\epsilon_w$
and $\sigma_m/\sigma_w$.

Equilibrium vesicle deformation can finally be calculated by
minimizing the total free energy (\ref{eq:equilibrium-requirement})
over $s_2$, yielding
\begin{equation}
  s_2 = \frac{1}{16} \frac{r_0^4 \epsilon_w E_0^2}{k_c}\, \left(
  \xi_\infty + \frac{\xi_1}{1 + \omega^2 \tau_1^2} + \frac{\xi_2}{1 +
  \omega^2 \tau_2^2} \right) \; ,
  \label{eq:deformation-omega}
\end{equation}
where $r_0^2 r_\mathrm{out}^2$ has been replaced by $r_0^4$,
consistent with the expansion in $E^2$ and retaining the terms with
the lowest energy. As we can see, the dependence of vesicle
deformation $s_2$ on the circular frequency $\omega=2\pi\nu$ contains
two dispersion terms of the Maxwell-Wagner type, which arise due to
the interfacial polarization and not due to intrinsic dispersion. It
is also worth emphasizing that the only approximation used in deriving
the expression (\ref{eq:deformation-omega}) is that of a small
deformation ($s_2 \ll r_0$).

The coefficients $\xi_\infty$, $\xi_1$, $\xi_2$, $\tau_1$ and $\tau_2$
figuring in (\ref{eq:deformation-omega}) can be simplified using
expansion in $(1 - \gamma)$, as $(1-\gamma)\lesssim 10^{-3}$ (the
vesicle radius is in the range 1--100~$\micro$m, while the thickness
of the phosphilipid membrane is approximately 4~nm). Expansion up to
the first order in $(1-\gamma)$ yields:
\begin{eqnarray}
  \fl \xi_\infty &= -\frac{2[-2\epar(\ew-\epar)^2 +
      (\ew+2\epar)\ew\Delta\epsilon +
      2\epar(\Delta\epsilon)^2]}{9\epar^2\ew}\, (1-\gamma) \; ,
      \label{eq:coefa} \\
  \fl \xi_1 &= \frac{-4\epar\sigma_m\sigma_w(\ew-\epar) + 2\sigma_w
    (\epar\sigma_w + \ew\sigma_m +
    2\epar\sigma_m)\Delta\epsilon}{9\epar^2\sigma_m^2}
    \left(\frac{\sigma_m}{\sigma_w} - \frac{\epar}{\ew}\right)
    (1-\gamma) \; , \label{eq:coefb} \\
  \fl \xi_2 &= \frac{8(\epar+\Delta\epsilon)}{9\epar}
  \left(\frac{\sigma_m}{\sigma_w} - \frac{\epar}{\ew}\right)(1-\gamma)
  \; , \label{eq:coefc} \\
  \fl  \tau_1 &= \frac{\epar}{\sigma_m} \; , \label{eq:tau1} \\
  \fl  \tau_2 &= \frac{\ew}{\sigma_w} \; . \label{eq:tau2}
\end{eqnarray}
We have introduced dielectric anisotropy $\Delta\epsilon =
\eperp-\epar$. The characteristic times $\tau_1$ and $\tau_2$
correspond to the trans-membrane and the trans-vesicle-interior
relaxations, respectively. It is worth noting that neither $\tau_1$
nor $\tau_2$ is affected by the anisotropy in this first-order
approximation.

\section{Results}

This section consists of four points. First, we plot the dependence of
deformation on frequency and comment on the effects of different
material constants. Next, we look for the conditions where an oblate
deformation ($s_2 < 0$) exists in the high-frequency limit. A similar
search is performed for the conditions where a prolate deformation
($s_2 > 0$) exists in the low-frequency limit. Finally, we plot and
analyse the dependence of the prolate-to-oblate transition frequency
on material constants.

\begin{figure}
  \includegraphics[width=10cm]{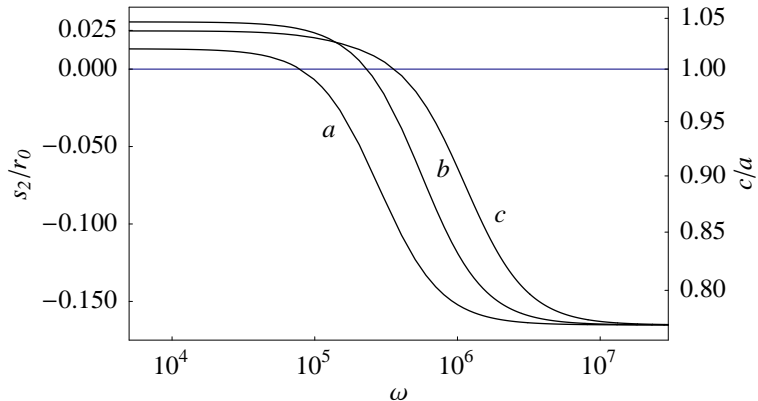}
  \caption{The theoretical dependence of vesicle deformation,
  expressed as either $s_2/r_0$ (left axis) or $c/a$ (right axis), on
  the circular frequency $\omega$ of the applied external electric
  field. The curves $a$, $b$ and $c$ correspond to the membrane
  conductivities of $0.5\times 10^{-5}$, $1.0\times 10^{-5}$, and
  $2.0\times 10^{-5}\;\mathrm{\Omega}^{-1}\textrm{m}^{-1}$,
  respectively. Other parameters used in the calculation were
  $r_0=10$~$\micro\textrm{m}$, $\gamma=0.9996$, $k_c=1.2\times
  10^{-19}\,\textrm{J}$, $E_0=100$~V/cm,
  $\sigma_w=10^{-4}\;\mathrm{\Omega}^{-1}\textrm{m}^{-1}$,
  $\epsilon_w=80\epsilon_0$, $\epar=2\epsilon_0$,
  $\Delta\epsilon=6\epsilon_0$.}
  \label{fig:4}
\end{figure}

\paragraph{Dependence of deformation on frequency.} 
Figure~\ref{fig:4} shows the vesicle deformation as a function of the
circular frequency $\omega$ of the applied external electric
field. Even though the expression for deformation
(\ref{eq:deformation-omega}) contains two dispersion terms, only one
of them is prominent on the diagram. This is because at values of the
parameters used for evaluation ($\epar\ll \epsilon_w$, $\sigma_m \ll
\sigma_w$), the magnitude of the second dispersion term is about two
orders of magnitude smaller, $\xi_2 \ll \xi_1$. This agrees with the
expressions (\ref{eq:coefb},\ref{eq:coefc}) where expansion in
$(1-\gamma)$ is used: $\xi_2 \propto \sigma_m/\sigma_w$, while $\xi_1
\propto (\epsilon_w/\epar)(\sigma_w/\sigma_m)$.

Figure~\ref{fig:vesicle}b provides a comparision of the results of the
model with experimental data. The parameters used to calculate the
dotted line in figure~\ref{fig:vesicle}b were $k_c=0.9\times
10^{-19}\,\textrm{J}$, $E_0=200$~V/cm,
$\sigma_w=10^{-4}\;\mathrm{\Omega}^{-1}\textrm{m}^{-1}$,
$\sigma_m=0.38\times 10^{-5}\;\mathrm{\Omega}^{-1}\textrm{m}^{-1}$,
$\epsilon_w=80\epsilon_0$, $\epar=2\epsilon_0$,
$\Delta\epsilon=5.2\epsilon_0$. While the results of the model
reproduce a general trend, it is clear that its agreement with the
experimental data is only qualitative. This can be attributed to the
perturbative nature of the model, which is only valid for small
deviations from a sphere only. This limitation is somewhat conflicting
with the experimental requirements, where significant deviations from
a sphere are clearly preferred for giving a more accurate readout.

The frequency-dependent prolate-to-oblate transition does not reflect
a change of sign in the forces of the electric field -- both in the
high- and the low-frequency limit, the forces induced by the electric
field pull the vesicle apart at poles and compress it around the
equator, thus favouring a prolate deformation. A lower energy for an
oblate shape at high frequencies stems from the shear term,
$f_\theta\,\delta{r}_\theta$. In the high-frequency limit, $f_\theta$
is zero for the isotropic case, and increases in magnitude with the
increasing anisotropy $\Delta\epsilon$.

\begin{figure}
  \includegraphics[width=10cm]{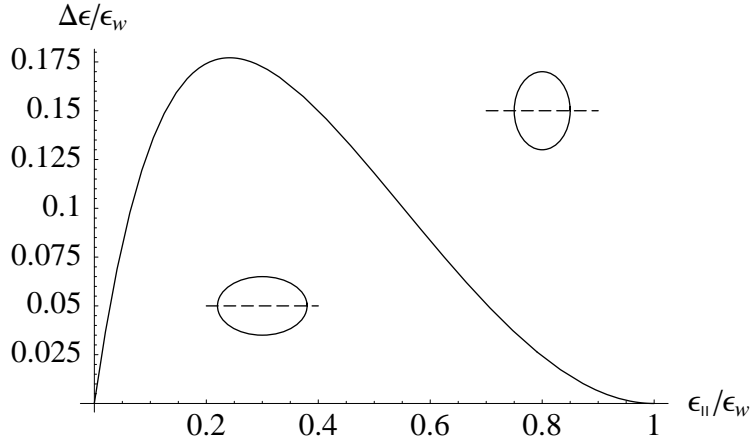}
  \caption{The boundary between the prolate and the oblate shape in the
  high-frequency limit as a function of the ratio between the
  permittivity of the membrane ($\epar$) and the permittivity of the
  aqueous solution ($\ew$). Deformation is oblate in the area above
  the curve and prolate in the area below the curve. For
  $\epar\ll\ew$, dependence can be approximated with a straight line
  $\Delta\epsilon/\ew = 2\epar/\ew$. The curve corresponds to a
  vesicle radius of 10~$\micro$m, yielding $\gamma=0.9996$.}
  \label{fig:anisotropy-graph}
\end{figure}

\paragraph{High-frequency limit.} 
High-frequency behaviour depends only on the dielectric permittivities
of the membrane and the aqueous solution, and is completely determined
by coefficient $\xi_\infty$ as can be seen from
(\ref{eq:deformation-omega}): if $\xi_\infty > 0$, the deformation is
prolate, and vice versa.  The sign of the exact expression for the
coefficient $\xi_\infty$ (without employing an expansion in
$(1-\gamma)$) depends on two parameters; here, they were chosen to be
$\epar/\epsilon_w$ and $\Delta\epsilon/\epsilon_w$. It is worth noting
that while the expression $\xi_\infty$ depends on $\gamma$ as well,
varying $\gamma$ can not change its sign. Furthermore, for all values
of vesicle radius which are of experimental interest, $\gamma \approx
1$ holds, so $\gamma$ does not significantly affect vesicle behaviour
in our model, therefore we did not put much emphasis on it.  Values of
$\epar/\epsilon_w$ and $\Delta\epsilon/\epsilon_w$ for which
$\xi_\infty=0$ holds can only be computed numerically. In
figure~\ref{fig:anisotropy-graph}, the transition boundary between
prolate and oblate shape in the high frequency limit on the
($\epar/\epsilon_w$, $\Delta\epsilon/\epsilon_w$) plane is plotted.

An insight into the high-frequency behaviour can also be obtained from
the expanded expression (\ref{eq:coefa}). If in addition to the
expansion in $(1-\gamma)$ an expansion in $\epar/\ew$ is employed (it
is estimated $\epar/\ew \sim 1/40$), one obtains the following
expression for the requirement for an oblate deformation in the
high-frequency limit:
\begin{equation}
  \Delta\epsilon \gtrsim 2\epar \; .
  \label{eq:HFcriterion}
\end{equation}

It can also be seen that in the isotropic case ($\Delta\epsilon = 0$),
coefficient $\xi_\infty$ attains the form (an expansion up to the
first order in $(1-\gamma)$ is used):
\begin{equation}
  \xi_\infty = \frac{4(\ew-\epar)^2}{9\epar\ew}\, (1-\gamma) \; ,
\end{equation}
which is always positive. Thus the deformation of a vesicle with
isotropic membrane permittivity is always prolate in the
high-frequency limit, which is consistent with the findings of
Winterhalter and Helfrich \cite{winterhalter:1988}.

\begin{figure}
  \includegraphics[width=10cm]{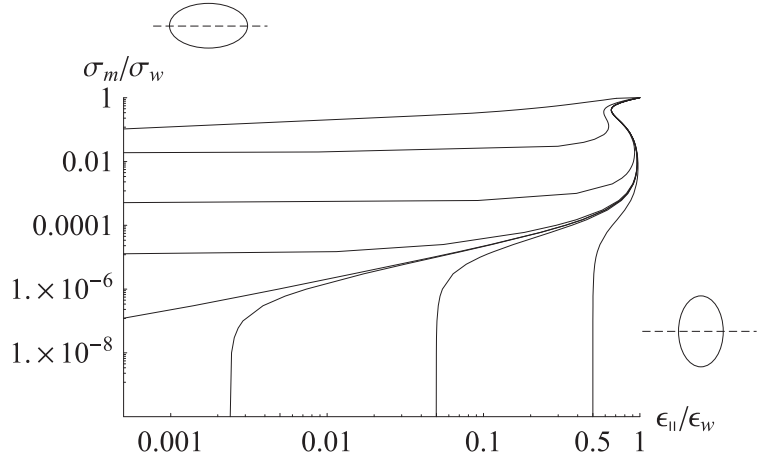}
  \caption{The boundary between the prolate and the oblate shape in
    the low-frequency limit as a function of the ratio between
    permittivity of the membrane $\epar$ and permittivity of water
    $\ew$ on the one hand and the ratio of membrane conductivity
    $\sigma_m$ and the conductivity of the aqueous medium $\sigma_w$
    on the other hand. Deformation is prolate in the area above the
    curve valid for a particular value of dielectric anisotropy and
    oblate in the area below it. The delineation lines correspond to
    different values of dielectric anisotropy (from bottom to top):
    $\Delta\epsilon$ = 0, 0.0144, 0.01592, 0.016, 0.018, 0.1, 3, and
    30 $\epsilon_0$. The curves are calculated for $\gamma=0.9996$.}
  \label{fig:conductivity-graph}
\end{figure}

\paragraph{Low-frequency limit.} 
The condition for prolate deformation in the low-frequency limit is
obtained from $\lim_{\omega\rightarrow 0}s_2$, yielding the criterion
$\xi_\infty + \xi_1 + \xi_2 = 0$ for the threshold between prolate and
oblate shapes.  Unlike the condition for the high-frequency limit, it
depends on the dielectric permittivities as well as the conductivities
of membrane and the aqueous medium. The sign of the exact expression
$\xi_\infty + \xi_1 + \xi_2$ depends on three parameters; here, they
have been chosen to be $\epar/\epsilon_w$,
$\Delta\epsilon/\epsilon_w$, and $\sigma_m/\sigma_w$. The comments on
the dependence of the expression on $\gamma$ in the high-frequency
limit apply here as well. Its zero can only be computed
numerically. In figure~\ref{fig:conductivity-graph}, the threshold
between prolate and oblate shape in the low frequency on the
($\epar/\epsilon_w$, $\sigma_m/\sigma_w$) plane is plotted for several
different values of $\Delta\epsilon$. The curves are plotted for
$\gamma=0.9996$, which corresponds to a vesicle radius of
10~$\micro$m.

It is worth noting that for small values of anisotropy
($\Delta\epsilon \approx 0$), curves intersect the abscissa. For
$\Delta\epsilon=0$, the intersection point is approximately $1/2 -
(1-\gamma)/3$, meaning that for ratios $\epar/\epsilon_w$ lower than
this value (this corresponds to all experimentally achievable cases),
the deformation of a vesicle with a membrane of negligible
conductivity in the low-frequency limit is prolate, while for the
cases $\epar/\epsilon_w > 1/2 - (1-\gamma)/3$, the deformation in the
low-frequency limit for the same vesicle is oblate. Increasing
dielectric anisotropy lowers the threshold value, and at
$\Delta\epsilon/\epsilon_w \approx 0.0002$, the deformation of a
vesicle with a non-conductive membrane is always oblate in the
low-frequency limit. This is consistent with the findings of
Winterhalter and Helfrich \cite{winterhalter:1988}, who have predicted
prolate deformation in the low-frequency limit for a vesicle with no
dielectric anisotropy and very small membrane conductivity
$\sigma_m/\sigma_w = 10^{-10}$.

Again, additional insight can be obtained when one employs expansions
in $(1-\gamma)$ and $\epar/\ew$, yielding the following requirement
for prolate deformation in the low-frequency limit:
\begin{equation}
  \sigma_m > \frac{\sigma_w}{2}\left( \frac{\epar-\eperp}{\epsilon_w}
  - 2\frac{\epar\eperp}{\epsilon_w^2}\right) \; .
  \label{eq:LFcriterion1}
\end{equation}
For $\epsilon_\perp = 3 \epsilon_\parallel$, which also fulfills the
requirement (\ref{eq:HFcriterion}) obtained for oblate deformation in
the high-frequency limit, (\ref{eq:LFcriterion1}) simplifies to
\begin{equation}
  \sigma_m \gtrapprox \frac{\epar}{\epsilon_w} \sigma_w \; .
  \label{eq:LFcriterion}
\end{equation}
In (\ref{eq:LFcriterion}), it has been taken into account that
$\epar,\eperp \ll \epsilon_w$. Prolate deformation in the
low-frequency limit for a vesicle with nonzero membrane anisotropy is
only possible when the ratio of membrane conductivity and the
conductivity of the aqueous solution exceeds the ratio of
permittivities of membrane and the aqueous solution.

Another special case is the isotropic case ($\epar = \eperp \ll
\epsilon_w$). Here, the expression on the right-hand side
(\ref{eq:LFcriterion1}) is negative, and so the inequality is always
fulfilled, meaning that the deformation of a vesicle with no
anisotropy in dielectric permittivity is always prolate in the
low-frequency limit.

\begin{figure}
  \includegraphics[width=10cm]{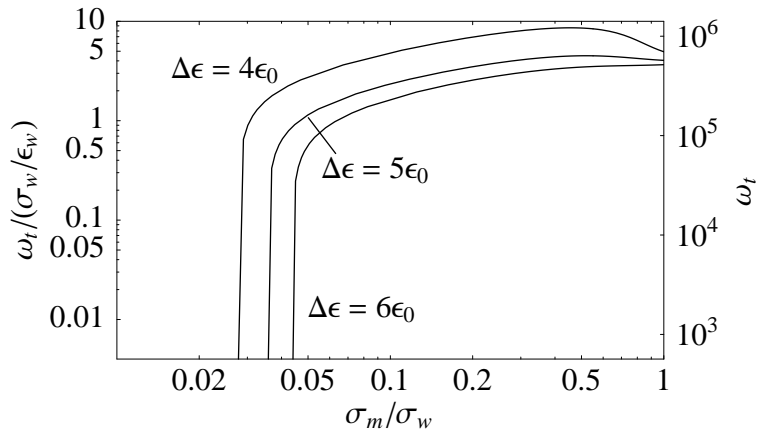}
  \caption{The dependence of the transition frequency between the
  prolate and the oblate shape on the value of the ratio between
  membrane conductivity and the conductivy of the aqueous medium
  ($\sigma_m/\sigma_w$), plotted for three different values of
  dielectric anisotropy ($\Delta\epsilon = 4,5,6\;\epsilon_0$). The
  dependence is calculated for $\epar=2\epsilon_0$, $\gamma=0.9996$.
  On the left axis, the scale is plotted in terms of dimensionless
  frequency $\omega_t/(\sigma_w/\ew)$; on the right axis, the scale is
  plotted in terms of circular frequency $\omega_t$ for
  $\sigma_w=10^{-4}\;\mathrm{\Omega}^{-1}\textrm{m}^{-1}$ and
  $\epsilon_w=80\epsilon_0$.}
  \label{fig:frekv-prehoda-prevodnost}
\end{figure}

\paragraph{Transition frequency.}
It is also of interest to plot the transition frequency between the
prolate and the oblate shape as a function of membrane
conductivity. This corresponds to finding the zeroes of the following
equation for different values of parameters:
\begin{equation}
  \xi_\infty + \frac{\xi_1}{1 + \omega^2 \tau_1^2} + \frac{\xi_2}{1 +
    \omega^2 \tau_2^2} = 0 \; .
\end{equation}
Parameter space can be reduced by using a dimensionless frequency, as
defined above. One can calculate the threshold dimensionless frequency
numerically. Its dependence on the ratio between membrane conductivity
and the conductivy of the aqueous medium ($\sigma_m/\sigma_w$) for
different values of dielectric anisotropy is plotted in
figure~\ref{fig:frekv-prehoda-prevodnost}. It can be seen that for
values of $\sigma_m/\sigma_w$ smaller than a certain threshold value
-- approximately given by (\ref{eq:LFcriterion}) -- one does not
obtain a prolate-to-oblate transition at all.

\section{Discussion}

The treatment presented in this paper largely follows the outline by
Winterhalter and Helfrich \cite{winterhalter:1988}. It is important to
note, however, that substituting isotropic values for the coefficients
$\xi_\infty$, $\xi_1$, $\xi_2$, $\tau_1$ and $\tau_2$ into expression
(\ref{eq:deformation-omega}) \emph{does not} lead to the same result
as equation~(21) in their article. The expansions in factors $\beta$
and $\gamma$ used by Winterhalter and Helfrich show non-analytical
behaviour, where consecutive expansions in $\gamma$ and $\beta$ yield
a result different from the one obtained by consecutive expansions
conducted in the opposite order, i.e., an expansion in $\beta$,
followed by an expansion in $\gamma$. Using Winterhalter and
Helfrich's values of material constants, one can see that $\beta >
\gamma$ holds in the high-frequency limit, while in the low-frequency
limit $\beta < \gamma$ holds. Due to this, the frequency dependence of
the simplified expression they obtained by series expansion differs
qualitatively from the exact expression from which the simplified
expression was derived -- while the deformation described by the exact
expression decreases with the increasing frequency, the simplified
expression increases. In this paper, we avoided using series expasion
in $\beta$ altogether, expressing it explicitely with the frequency
$\omega$ and the material constants independent of frequency. The
expression (\ref{eq:deformation-omega}) may provide a better
approximation to the exact expression than the one used by
Winterhalter and Helfrich, as it retains the frequency dependence of
the exact expression, while the expansion they used employs an
expansion in $\beta$ and is thus actually limited to the low-frequency
range only.

In the previous section, we have developed criteria within the
framework of our model, under which the vesicle shape is prolate in
the low-frequency limit and oblate in the high-frequency limit. The
criteria we obtained however opened a couple of questions about their
physical relevance: the tangential component of permittivity is
predicted to exceed its normal component threefold, the membrane
conductivity is predicted to be higher than the earlier estimates
\cite{winterhalter:1988}.
\begin{enumerate}
\item\emph{Membrane permittivity.}  Within the framework of the
presented model, a criterion for an oblate shape in the high-frequency
limit was obtained, requiring that the tangential component of the
dielectric permittivity tensor of the phospholipid membrane exceeds
its radial component by approximately threefold: $\eperp \gtrapprox
3\epar$. This is in qualitative agreement with an abundance of
experimental evidence claiming that the zwitterionic
phosphatidylcholine headgroup is oriented almost parallel to the
surface of a hydrated bilayer and its movement is essentially confined
to a plane normal to the membrane, employing techniques such as X-ray
diffraction, neutron diffraction, ${}^2$H nuclear magnetic resonance
(NMR) and ${}^{31}$P NMR as well as molecular simulation studies (see
\cite{Raudino:2001} and the references therein).

\item\emph{Membrane conductivity.}  Another criterion was obtained for
a prolate vesicle shape in the low-frequency limit, which required
that the ratio of membrane conductivity and conductivity exceeds the
ratio of membrane permittivity and the permittivity of the aqueous
medium. In practice, this means a relatively high value for membrane
conductivity: $\sigma_m \sim 10^{-5}\:\textrm{S\,m}^{-1}$. This
exceeds not only the value for a thin layer of oil, $\sigma_m \sim
10^{-14}\:\textrm{S\,m}^{-1}$ \cite{winterhalter:1988}, but also some
other estimates for membrane conductivity ($\sigma_m \sim
10^{-7}\:\textrm{S\,m}^{-1}$) \cite{Sukhorukov:2001}. The high value
of conductivity could be attributed to the impurities in the lipid and
the transient submicroscopic pores in the membrane (step 1 in the
accepted five-step description of electropermeabilization,
\cite{Teissie:2005}). A simple model treating the intact membrane and
the pores as resistors connected in parallel shows that the total area
of pores must account for approximately 1/1000 of the total membrane
area, which corresponds to 1000 pores with an area of
100~$\textrm{\AA}^2$ per 1~$\micro\textrm{m}^2$ of membrane area. This
significantly -- $10^6$--$10^8\times$ -- exceeds the density of
spontaneous pore formation in the membrane \cite{Raphael:2001}. This
latter value increases, however, with an the applied electric field
\cite{Neu:1999}, even though the electric field strength in our setup
is approximately 70--$300\times$ lower than its threshold value
neccessary for the pore expansion into micrometer-sized ``macropores''
\cite{Riske:2005}. An initial membrane tension present in a vesicle
with a relative volume very close to 1 (the estimate for the vesicle
shown in figure~\ref{fig:vesicle} is $v=0.9993\pm0.0007$) also works
in the same direction. Altogether, a high value of membrane
conductivity predicted by the presented model remains a poorly
understood effect, and awaits further investigations before a
definitive conclusion can be made.
\end{enumerate}

\paragraph{Geometry considerations.}
It has been our intention to build the simplest possible model which
would explain the observed frequency-dependent prolate-to-oblate shape
transition for small deformations and yield a clear physical picture
of the phenomenon. The validity of the model decreases once we depart
from small deformation. A possible extension of the model lies in a
more appropriate description of the vesicle geometry, i.e. spheroidal
instead of spherical. An effort in the suggested direction might
benefit from the treatment of a spheroid vesicle with a permeable
membrane \cite{Sokirko:1994}, as well as from the calculations of
transmembrane voltage for the case of zero membrane conductance
\cite{Kotnik:2000b,Gimsa:2001f} and the solution for the electric
potential for the general anisotropic case \cite{Ambjornsson:2003}.

\section{Conclusions}

The proposed model provides a theoretical explanation for the observed
prolate-to-oblate transition of phosholipid vesicle shape, when the
frequency of the applied AC electric field increases, in the case when
the electrical properties of the aqueous solution inside the vesicle
do not differ from the properties of the aqueous solution outside the
vesicle. Vesicle deformation at varying frequency was calculated
numerically. Using an expansion into Taylor series we also derived a
simplified expression for the deformation. The expression retains the
frequency dependence of the exact expression and may provide a better
substitute for the expression that Winterhalter and Helfrich obtained
by series expansion, which was found to be valid only in the limit of
low frequencies. The model with the anisotropic membrane permittivity
imposes two constraints on the values of material constants:
tangential component of dielectric permittivity tensor of the
phospholipid membrane must exceed its radial component by
approximately a factor of 3; and membrane conductivity has to be
relatively high, approximately one tenth of the conductivity of the
external aqueous medium. Both constraints seem to be justified to a
certain degree by the claims found in the literature.

\ack{This work has been supported by the Slovenian Research Agency
research grant P1-0055.}

\section*{References}
\bibliographystyle{iopart-num}
\bibliography{prolate-oblate}

\end{document}